\documentclass{pasj00}
\draft

\begin{document}
\SetRunningHead{Author(s) in page-head}{Running Head}

\title{Suzaku Observation of the Symbiotic X-Ray Binary IGR J16194--2810}

\author{Yuiko \textsc{Kitamura}, Hiromitsu \textsc{Takahashi}}
\and \author{Yasushi \textsc{Fukazawa}}
\affil{Department of Physical Science, School of Science, Hiroshima University, 1-3-1 Kagamiyama, Higashi-Hiroshima 739-8526}
\email{kitamura@hep01.hepl.hiroshima-u.ac.jp, hirotaka@hep01.hepl.hiroshima-u.ac.jp, fukazawa@hep01.hepl.hiroshima-u.ac.jp}

%


\maketitle

\begin{abstract}

We observed IGR J16194--2810 in the low/hard state with the Suzaku X-ray satellite in 2009.
The source is a Symbiotic X-ray Binary (SyXB) classified as a category of a Low-Mass X-ray Binary (LMXB), since the system is composed of an M-type giant and
probably a neutron star (NS).
We detected the 0.8--50 keV signal with the XIS and HXD-PIN.
The 2--10 keV luminosity was $L \sim 7 \times 10^{34}$ erg s$^{-1}$ corresponding to $\sim 10^{-3} L_{\rm Edd}$, where $L_{\rm Edd}$ is the Eddington Luminosity of a 1.4 $M_{\odot}$ NS and a source distance of 3.7 kpc is assumed.
The luminosity is similar to those of past observations.
The spectral analysis showed that there are two emission components below and above $\sim$ 2 keV.
The hard emission component is represented by a Comptonized black-body emission model with the seed-photon temperature $\sim$ 1.0 keV and the emission radius
 $\sim$ 700 m.
The seed photon is considered to come from a small fraction of the NS surface.
The soft component is reproduced by either a raw black-body ($\sim$ 0.4 keV, $\sim$ 1.7 km) or a Comptonized emission ($\sim$ 0.1 keV, $\sim$ 75 km).
We think the origin is the emission from other part of the NS surface or the accreting stream.
The physical parameters of the hard emission component of IGR J16194--2810 are compared with those of an SyXB (4U 1700+24) and LMXBs
(Aql X-1 and 4U 0614+091).
This comparison reveals that these SyXBs in the low/hard state have a smaller radiation region ($<$ 1 km) on the NS surface with a higher seed-photon temperature ($\sim$ 1 keV) than the compared LMXBs.
\end{abstract}

\section{Introduction}\label{sec:intro}
X-ray binaries are classified into Low-Mass X-ray binaries (LMXBs) and High-Mass X-ray binaries (HMXBs).
187 LMXBs have been discovered in our Galaxy \citep{liu_lmxb}.
LMXBs are binaries consisting of a low-mass companion star and a main compact star (a black hole or a weakly magnetized neutron star (NS)), and thus they are old systems.
HMXBs are binaries whose companion is a high-mass star.

Symbiotic X-ray binaries (SyXBs) are the system of a compact star with a red giant, and previously defined as LMXBs \citep{masetti_syxb}.
\citet{masetti_syxb2} and \citet{smith_syxb}
 detected 6 SyXBs.
In addition, 
\citet{lu_syxb}
 newly listed 5 candidate systems of SyXBs,
although 1RXJ 180431.1--273932 and 2XMM J174016.0--290337 were reported as cataclysmic variables in
 Masetti \etal\ (2012ab),
and IGR J16393--4643 was identified as a HMXB \citep{corbet_igrj16393,bodaghee_igrj16393}.
Comparing SyXBs with general LMXBs, one of the remarkable features is
long NS spin periods; for example,
GX 1+4 and GX 1954+31 have 110--157 s (e.g., \cite{chakrabarty_gx1+4}, \cite{gonzalez_gx1+4}) and $\sim$ 5 h \citep{corbet_4u1954}, respectively.
Therefore, understanding the evolution of NSs and LMXBs, detailed investigation of SyXBs is very important.

When LMXBs have the luminosity of 10$^{36}$ erg s$^{-1}$ in the high/soft state, Compton corona is suggested to exist around the NS and the X-ray spectrum is often described by the soft and hard components; the former component is Disk black body (DBB) and the latter is Comptonization of black body (BB) from the NS surface \citep{mitsuda_hard}.
As the X-ray luminosity decreases below 10$^{36}$ erg s$^{-1}$, LMXBs become
the low/hard state in which the X-ray spectrum becomes harder
(e.g., \cite{barret_hard}, \cite{lin_hard}, \cite{sakurai_aqlx1}).
SyXBs have been often observed in the low/hard state.
However, the X-ray spectrum of SyXBs has not been systematically studied, and thus the origin of X-ray radiation is not fully understood.
The purpose of this study is to explore the X-ray spectrum and variability of SyXBs in the low/hard state in terms of the following points;
how the X-ray spectral features of SyXB are,
and how the temperature and optical depth of Compton cloud around 
the compact star (most cases are the NS) and
accreting stream
are.

4U 1700+24 is one of NS-SyXBs and it is nearest from the earth among NS-SyXBs; the distance is $\sim$ 420 pc.
This object was observed with XMM-Newton and Suzaku when it was in the low/hard state with a luminosity of 10$^{-6}$--10$^{-4}L_{\rm Edd}$ (0.3--10 keV)
\citep{masetti_4u1700,tiengo_4u1700,nagae_4u1700}.
\citet{nagae_4u1700}
 reported that the X-ray spectrum was
reproduced by the combination of the inverse Compton of DBB for the
soft component and that of BB for the hard component, together with some
low-energy lines.

Here, we observed an SyXB IGR J16194--2810 with Suzaku.
This object consists of a M2 III star and
 probably a NS,
and the distance from the earth is $\sim$ 3.7 kpc.
It was first observed in the hard X-ray band above 20 keV with INTEGRAL on 2nd IBIS survey
(\cite{bird_ibis2nd}; see also the 3rd IBIS survey of \cite{bird_ibis3rd})
 and IBIS extragalactic survey, and the flux was reported to be $\sim$ 3$\times$10$^{-11}$ erg cm$^{-2}$ s$^{-1}$ in 20--100 keV \citep{bassani_ibis}.
This source was also observed with the X-ray Telescope (XRT, 0.3--10 keV;
\cite{burrows_xrt})
 on Swift, and the flux was $\sim$ 4.4$\times$10$^{-11}$ erg cm$^{-2}$ s$^{-1}$ in 2--10 keV \citep{masetti_syxb2}.
Therefore, the 2--10 keV luminosity is $L \sim 7.2 \times 10^{34}$ erg s$^{-1}$.
Since the observational dates and exposures of the INTEGRAL and Swift observations are different,
in this paper, we report the results of the 0.8--50 keV wide-band and simultaneous observation by Suzaku and discuss the physical state of this object by comparing with an SyXB 4U 1700+24 and typical LMXBs such as
Aql X-1 and 4U 0614+091.

\section{Observation and Data Reduction}\label{sec:obs}
IGR J16194--2810 was observed with Suzaku on February 5--6, 2009. The
net exposure time was about 46 ks. Suzaku has four X-ray CCD cameras
called the X-ray Imaging Spectrometer (XIS-0, XIS-1, XIS-2, and XIS-3)
which are sensitive in 0.2--12 keV.
The XIS-0, XIS-2 and XIS-3 are front-illumination (FI) CCDs and while the XIS-1
is back-illumination (BI) one \citep{koyama_xis}.
Due to the damage, the XIS-2 has not been operated since November 9, 2006.
Suzaku has a non-imaging instrument called the Hard X-ray Detector (HXD)
\citep{takahashi_hxd,kokubun_hxd}.
The HXD has two types of detectors, the Si-PIN photo diodes (PIN) and the
GSO scintillators, which are sensitive in 10--600 keV.

IGR J16194--2810 was observed at the HXD nominal position.
The XIS was operated in the $3\times3$ and $5\times5$ editing modes and the
normal clocking mode.
The signal was detected in 0.8--50 keV with the XIS and HXD-PIN.
For the spectral analysis, we chose Good Time Intervals (GTIs) when both XIS and HXD-PIN were active, and obtained the exposure of 36 ks.
The observed source count rates after subtracting the background were 1.44 and 1.51 counts s$^{-1}$ in 0.8--10 keV by XIS-FI and XIS-BI, respectively, and 0.03 counts s$^{-1}$ in 20--50 keV by HXD-PIN.

\subsection{XIS Reduction}

We analyzed the cleaned event file version 2.3 by using analysis software package HEASOFT 6.10 and the calibration database CALDB latest on 2011 April 1.
We defined the source region as a circle with a radius of 2$'$.7 from the source position, which we determine by looking at the XIS image.
The background region is taken as an annulus with a radius of 3$'$.3--5$'$.0 from the source position.
We generated the redistribution matrix files (RMFs) and ancillary response files (ARFs) by xisrmfgen and xissimarfgen of FTOOLS.
We added the XIS-0 and XIS-3 spectra by mathpha of FTOOLS.
Also, we multiplied RMF and ARF by using marfrmf of FTOOLS per each detector and we used these files as the RSP file.
We added the XIS-0 and XIS-3 RSP files by addrmf of FTOOLS.

\subsection{HXD Reduction}

We extracted the HXD-PIN spectrum from the cleaned event file version 2.3 by using hxdpinxbpi of FTOOLS,
where contribution of Cosmic X-ray background is subtracted assuming the spectral shape of exponential cut-off power law (photon index = 1.29, cut-off energy = 40 keV, flux at 1 keV = 9.412 $\times 10^{-3}$ photons keV$^{-1}$ cm$^{-2}$ s$^{-1}$ FOV$^{-1}$) \citep{boldt_cxb}.
For the non X-ray background (NXB), we used the tuned NXB file \citep{fukazawa_bgd},
 and the response file ae\_hxd\_pinhxnome5\_20080716.rsp was used.

During this observation, the HXD temperature was high and the lower energy band of the HXD-PIN suffered from noise events.
Therefore, we analyzed the HXD-PIN data above 20 keV when we fitted the spectra.
The cross-normalization factor of 1.18 is introduced to explain the absolute flux difference to the XIS
(http://heasarc.gsfc.nasa.gov/docs/suzaku/analysis/abc/abc.html).

\section{Analysis}\label{sec:ana}
\subsection{Analysis of the Time Average Spectra}\label{sec:ana_ave}

\begin{figure}
  \begin{center}
    \FigureFile(80mm,80mm){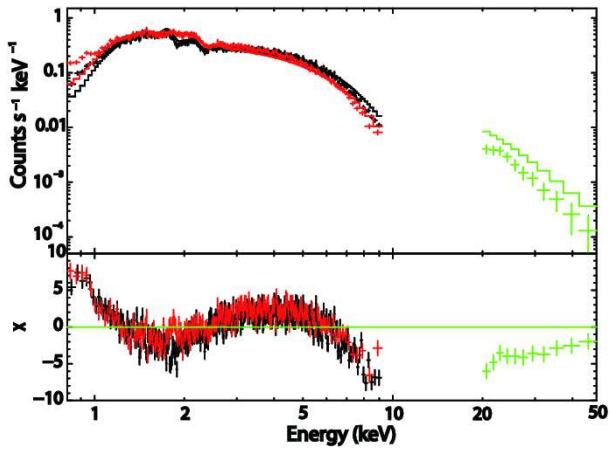}
  \end{center}
  \caption{Spectra fitted with the model~1 "phabs$\times$power-law". Black, red, and green data
 correspond to XIS-FI, XIS-BI, and HXD-PIN, respectively.
 A solid line represents the total model. The bottom panel shows a residual.}
\label{fig1}
 \end{figure}

We analyzed the XIS-FI, BI and HXD-PIN spectra by using XSPEC version 12.7.0.
Hereafter, the error is estimated as a 90\% confidence range ($\Delta \chi^{2}$ = 2.7) for one parameter.
First of all, we tried to fit the spectra with the model~1 "phabs$\times$power-law".
In this model, "phabs" represents the Galactic interstellar absorption with the cross section data of \citet{church_abs}.
Free parameters are the absorption column density $N^{\rm Gal}_{\rm H}$,
 the photon index $\Gamma$,
and the normalization of the power-law.
We show the fitting result of the model~1 in table~\ref{tab1} and figure~\ref{fig1}.
This model does not reproduce the observed spectra with a reduced $\chi^2_{\nu}$ of 5.35 (d.o.f. = 591).
Positive residuals are seen below 1 keV and at 2--6 keV, and a large negative residual appears above 6 keV.
To represent the high-energy negative residual, we applied "phabs$\times$compTT" which is the same model by \citet{masetti_syxb2}.
The compTT component models Comptonized radiation of BB
 seed photons \citep{titarchuk_comptt}.
As the result, the model can reproduce the spectra above 10 keV with the parameters of the seed-photon BB temperature of 0.74 keV and the optical depth of Compton cloud of $\sim$ 0.1.
However, the low-energy residual still remains, and $\chi^{2}_{\nu}$ is 1.48 (d.o.f = 589) thus not formally acceptable.

Second, we fitted the spectra with a conventional model for the low/hard state of LMXBs; "phabs(diskbb + compPS (seed=BB))" (model~2).
The diskbb (DBB) model represents Multi-Color Disk blackbody
(e.g., \cite{mitsuda_mcd}, \cite{makishima_mcd}).
Parameters are the inner-radius temperature of the accretion disk $T_{\rm DBB}$ and Normalization$_{\rm DBB}$.
In the DBB (and the following compPS (seed=DBB)) model, the obtained Normalization$_{\rm DBB}$ can be converted into the physical value of the disk inner radius $R_{\rm DBB}$ with Normalization$_{\rm DBB}$ = ($R_{\rm DBB}$ (km))$^2$ / (0.37
 (10 kpc))$^2 {\rm cos}\theta$, where $\theta$ is an inclination angle of the disk and the source distance of 3.7 kpc is assumed.
Since $\theta$ of this source is uncertain, we assume at 60$^{\circ}$ considering the fact that the eclipse has not been observed for this object.
If $\theta$ is measured as 0$^{\circ}$ or 85$^{\circ}$, the $R_{\rm DBB}$ value in this paper should be multiplied by a factor of 0.5 or 5.7
(other parameters such as the temperature and optical depth of the Compton cloud are independent from $\theta$ in the DBB and compPS (seed=DBB) models).
The compPS model \citep{poutanen_compps}
 is physically the same as the above compTT, and it can estimate the radiation radius of the seed photon.
Free parameters are BB temperature $T_{\rm BB}$, the optical depth of Compton cloud $\tau_{\rm BB}$ and Normalization$_{\rm BB}$.
Since the Comptonizing electron temperature $T_{\rm e}$ cannot be determined from the observed spectra, it is fixed to 100 keV; in fact, this value represents the lower limit of $>$ 90 keV.
As summarized in table~\ref{tab2},
 the model~2 well reproduced the observed spectra with a reduced $\chi^2_{\nu}$ of 1.13 (d.o.f. = 588).
The improvement from the model~1 (the $\chi^2_{\nu}$ difference with $F$-statistic value of 183) is significant more than the 3$\sigma$ level.
However, the inner radius of the accretion disk becomes $R_{\rm DBB} \; \sim 0.9 \; (\leq \; 1.5)$ km assuming the distance of 3.7 kpc and the inclination angle of 60$^{\circ}$.
This is not likely since it is smaller than the NS radius.

Instead of the DBB model, we next tried Astrophysical Plasma Emission Code (APEC) model which represents the emission from collisionally-ionized diffuse gas (http://atomdb.org/),
since the accreting stream originally from the stellar wind of the red giant may keep hot diffuse gas rather than forming the optically-thick disk.
As shown in table~\ref{tab3}, this model "phabs(APEC + compPS (seed=BB))" can also represent the spectral shape with a reduced $\chi^2_{\nu}$ of 1.14 (d.o.f. = 588) apparently.
However, due to the lack of Fe-L emission features around 1 keV,
the spectral continuum below 2 keV requires too low abundance of $<$ 0.004 solar value or too high temperature $\sim$ 40 keV when the abundance is fixed at 0.3 solar value.
Therefore, we think that the model~2 with either of DBB or APEC model cannot be physically accepted.

\begin{figure}
  \begin{center}
    \FigureFile(80mm,80mm){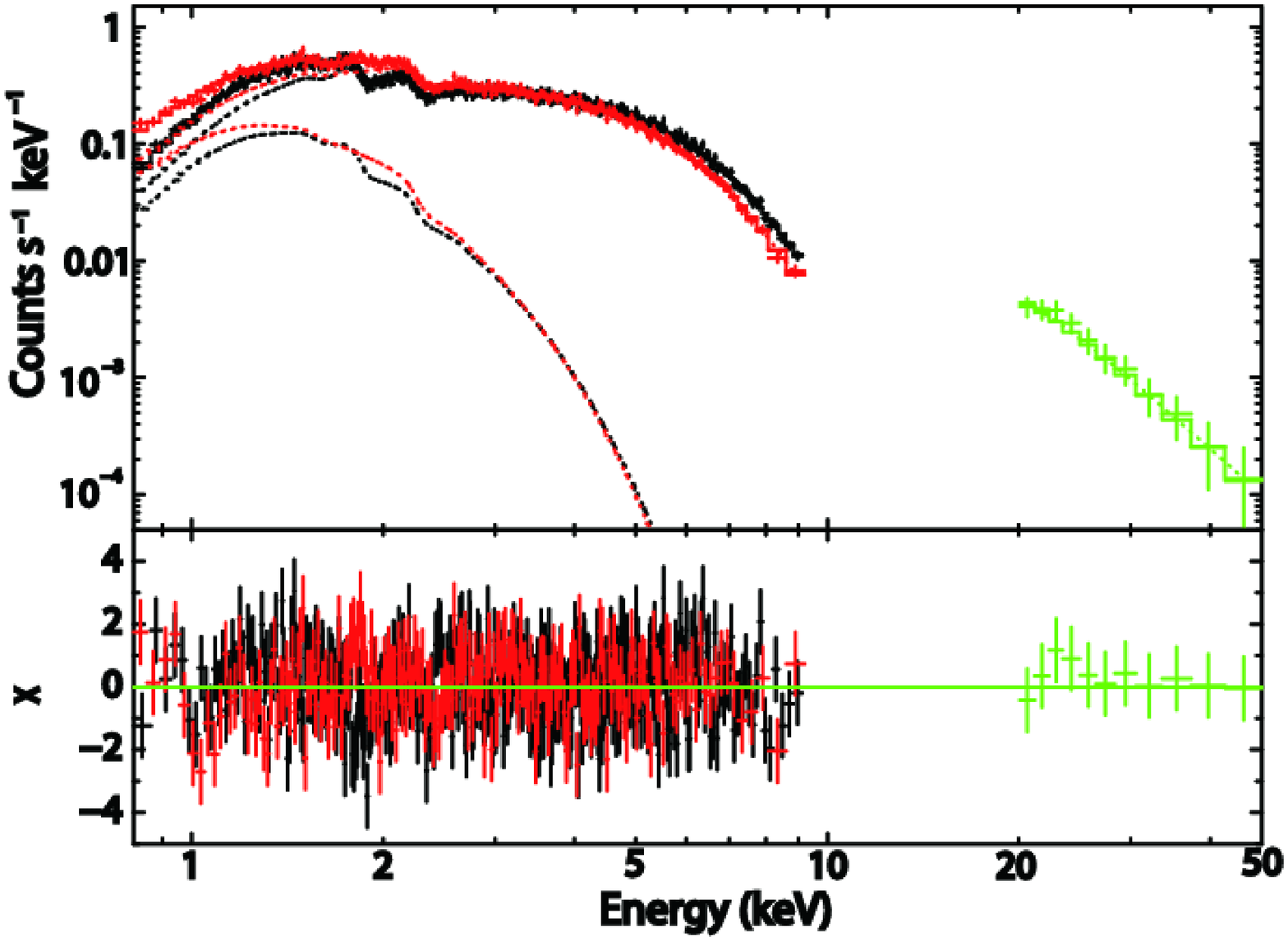}
    \FigureFile(80mm,80mm){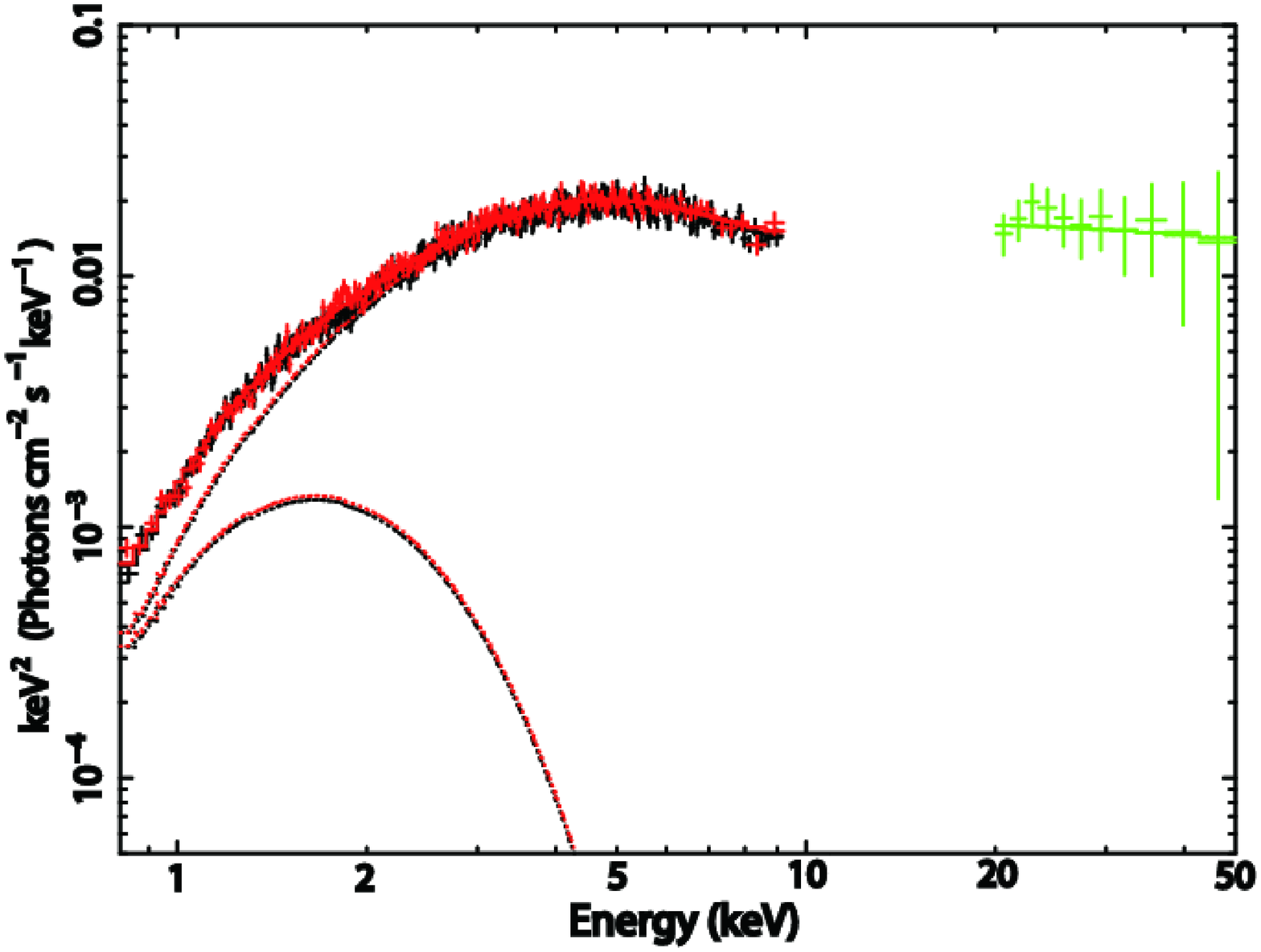}
 \end{center}
  \caption{(Left) the same as figure~1 but with the model~2 "phabs(BB + compPS (seed=BB))". Dash lines represent each model component.
(Right) the deconvolved energy spectrum of the model~2 (BB).
}
\label{fig2}
\end{figure}

Since the unrealistically small emission radius of the DBB may suggest the soft emission component comes from the NS surface,
we applied the model~2 of "phabs(BB + compPS (seed=BB))".
As summarized in table~\ref{tab2} and figure~\ref{fig2}, the model can reproduce the spectra
with a reduced $\chi^2_{\nu}$ of 1.14 (d.o.f. = 588),
and the obtained parameters are physically reasonable; the BB radius is $\sim$ 1.7 km with the temperature of $\sim$ 0.4 keV.

\begin{figure}
  \begin{center}
    \FigureFile(80mm,80mm){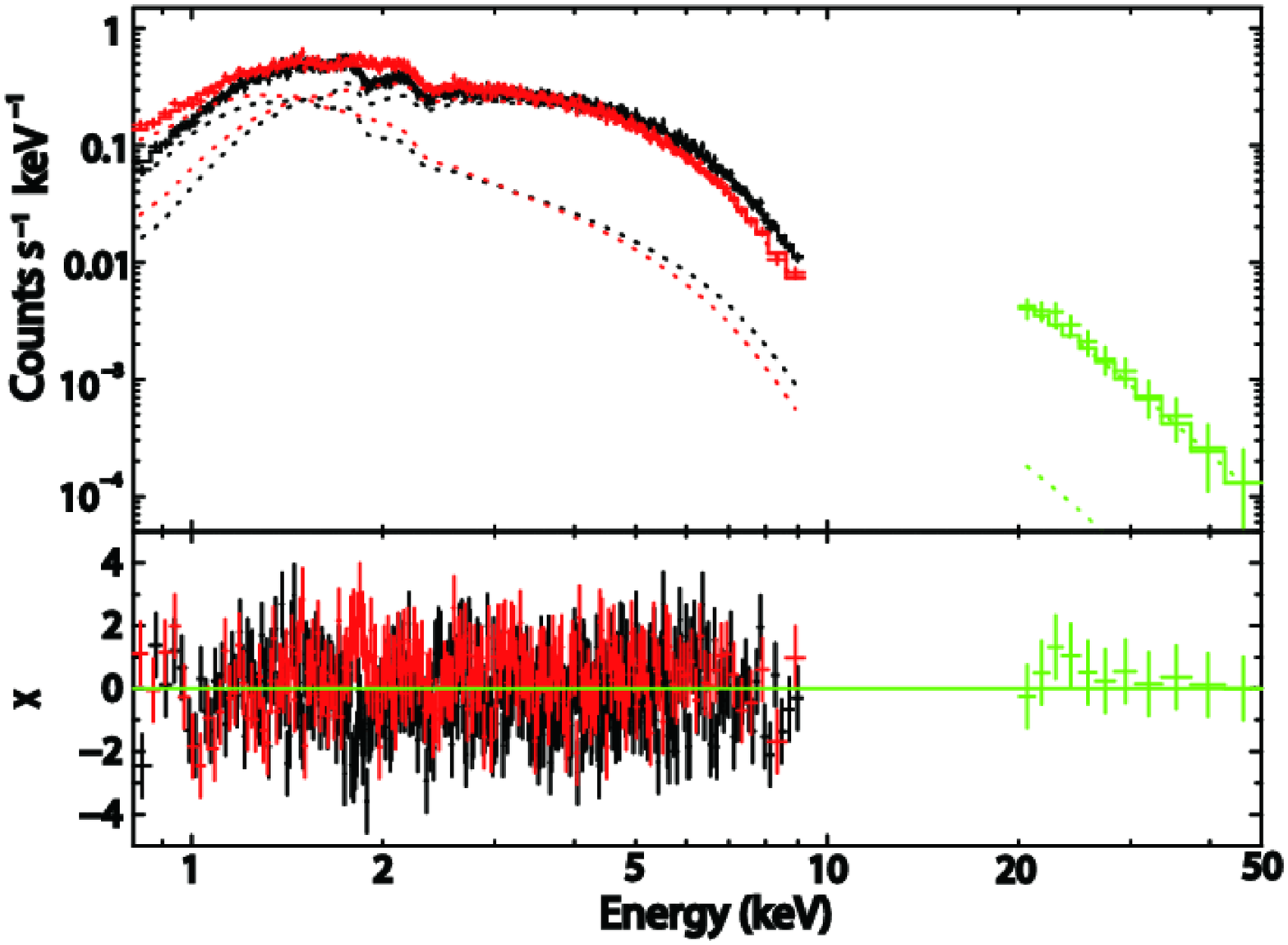}
    \FigureFile(80mm,80mm){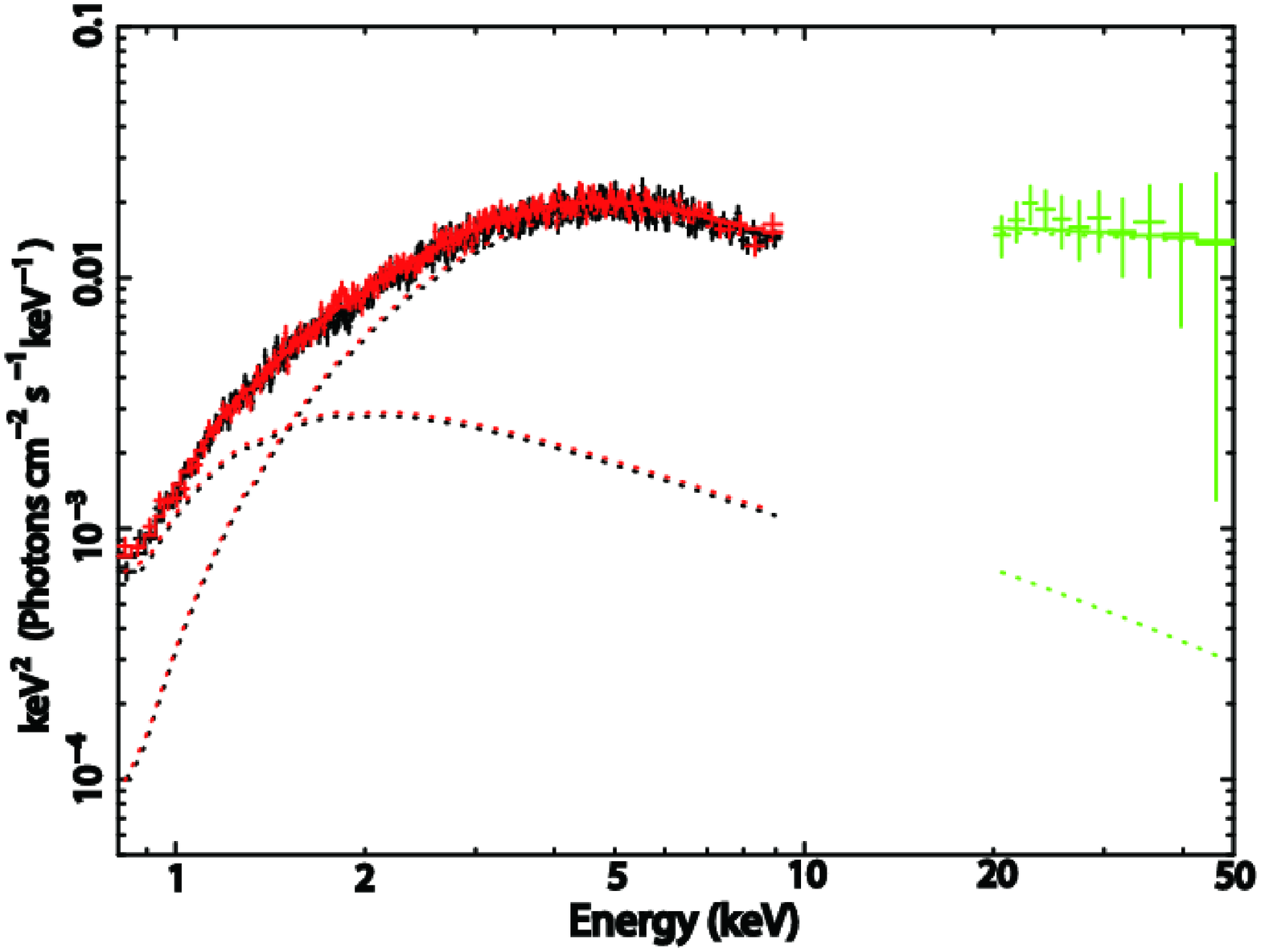}
  \end{center}
  \caption{(Left) the same as figure~2 but with the model~3 "phabs(compPS (seed=DBB) + compPS (seed=BB))". (Right) the deconvolved energy spectrum of the model~3.}
\label{fig3}
\end{figure}

Overestimation of the disk temperature may be another possibility to fake the too small disk inner radius.
The disk temperature would be much lower and the DBB photons would be
upscattered by Comptonization into the soft X-ray band.
Then, we also fitted the spectra with the model~3 "phabs(compPS (seed=DBB) + compPS (seed=BB))", where the DBB in the model~2 is replaced by the compPS (seed=DBB).
The shape of the Compton cloud and the electron temperature $T_{\rm e}$ are assumed to be spherical and 100 keV, respectively, the same as the compPS (seed=BB).
The model~3 can reproduce the observed model as well as the model~2 with a reduced $\chi^{2}_{\nu}$ of 1.14 (d.o.f = 587),
 as shown in table~\ref{tab4} and figure~\ref{fig3}.
In this model, the parameters of the seed DBB cannot be well constrained.
When $T_{\rm DBB} \; \sim$ 0.1 keV, $T_{\rm DBB}$ could be constrained from the spectral curvature below 1 keV, but the interstellar absorption prevents us to constrain it.
Nevertheless, the error region contains the parameter set that the disk temperature is low, the optical depth of Compton cloud is large, and the disk inner radius is reasonable to be $>$ 10 km.
Therefore, this model gives a possible solution of the spectral modeling.
Note that the origin of the seed photons with the temperature of $\sim$ 0.1 keV is not necessarily the multi-color disk black-body emission,
since the spectral shape of the seed-photon emission is not well determined.
On the other hand, the BB temperature for compPS (seed=BB) is determined by the apparent concave shape of the spectra around 10 keV.
Therefore, the uncertainty of $T_{\rm BB}$ and $R_{\rm BB}$ is smaller unlike the case of compPS (seed=DBB).

We concluded that the model~2 (BB) or the model 3 is physically possible to reproduce the time-average spectra.
The observed 2--10 keV flux and luminosity described in table~\ref{tab2}, \ref{tab3} and \ref{tab4} are $4 \times 10^{-11}$ erg s$^{-1}$ and $7 \times 10^{34}$ erg s$^{-1}$, respectively,
which are comparable to those of the previous observations \citep{masetti_syxb2}.

\begin{table}
\begin{center}
\caption{Fitting results with the model~1.}
\label{tab1}
\begin{tabular}{llc} \hline \hline

Model & Parameter &  \\ \hline
phabs & $N_{\rm H}$$^{\rm Gal}$ ($\times$10 $^{22}$cm$^{-2}$) & 0.97 \\
power-law & $\Gamma$ & 1.75 \\
 & Normalization\footnotemark[$*$] &  1.22$\times10^{-2}$\\ \hline
 &$\chi^{2}_{\nu}$ (d.o.f) & 5.35 (591) \\ \hline

\multicolumn{3}{@{}l@{}}{\hbox to 0pt{\parbox{85mm}{\footnotesize
  \par\noindent
  \footnotemark[$*$] Normalization = photons keV$^{-1}$ cm$^{-2}$ s$^{-1}$ at 1 keV
 }\hss}}
\end{tabular}
\end{center}
\end{table}

\begin{table}
\begin{center}
\caption{Fitting results with the model~2 of diskbb (DBB) and BB.}
\label{tab2}
\begin{tabular}{llcc} \hline \hline

Model & Parameter &  diskbb & BB \\ \hline
phabs & $N_{\rm H}$$^{\rm Gal}$ ($\times$10 $^{22}$cm$^{-2}$) & 0.40$^{+0.07}_{-0.04}$ & 0.33$^{+0.03}_{-0.04}$\\
diskbb & $T_{\rm DBB}$ (keV) & 0.60$^{+1.13}_{-0.13}$ & \\
 & Normalization$_{\rm DBB}$\footnotemark[$*$] & 3.1($\leq$ 8.0) & \\
 &$R_{\rm DBB}$ (km) & 0.9($\leq$ 1.5) & \\
BB & $T_{\rm BB}$ (keV) & & 0.36$^{+0.04}_{-0.03}$ \\
 & Normalization$_{\rm BB}$\footnotemark[$\dagger$] & & $4.2^{+0.9}_{-0.4} \times 10^{-5}$ \\
 &$R_{\rm BB}$ (km) & & 1.7$\pm$0.4 \\
compPS(seed=BB) &$T_{\rm BB}$ (keV) & 1.05$^{+0.07}_{-0.04}$ & 1.02$^{+0.04}_{-0.03}$\\
 &$\tau_{\rm BB}$ & 0.56$^{+0.04}_{-0.05}$ & 0.58$^{+0.03}_{-0.04}$ \\
 &Normalization$_{\rm BB}$\footnotemark[$\ddagger$] & 3.4$^{+0.6}_{-1.9}$ & 3.9$^{+0.3}_{-0.7}$\\
 &$R_{\rm BB}$ (m) & 680$^{+60}_{-220}$ & 730$^{+30}_{-70}$ \\ \hline
 &$\chi^{2}_{\nu}$ (d.o.f) & 1.13 (588) & 1.14 (588) \\
 & $F_{2 - 10}$ \footnotemark[$\S$] & 4.08 & 4.08 \\
 & $L_{2 -10}$ \footnotemark[$\|$] & 6.92 & 6.88 \\ \hline

\multicolumn{4}{@{}l@{}}{\hbox to 0pt{\parbox{85mm}{\footnotesize
  \par\noindent
  \footnotemark[$*$] Normalization = ($R_{\rm DBB}$ (km))$^2$ / (0.37
 (10 kpc))$^2 {\rm cos}\theta$, where $R_{\rm DBB}$ is an inner radius of
the disk in a unit of km. $\theta$ is an inclination angle and 60$^{\circ}$ is assumed.

  \footnotemark[$\dagger$] Normalization = $L_{\rm 39}$ /
 (0.37 (10 kpc))$^2$, where $L_{\rm 39}$ is the BB luminosity in a unit of 10$^{39}$ erg s$^{-1}$.

  \footnotemark[$\ddagger$] Normalization = ($R_{\rm BB}$ (km))$^2$ /
 (0.37 (10 kpc))$^2$, where $R_{\rm BB}$ is a radius of the BB seed-photon emission in a unit of km.

  \footnotemark[$\S$] 2--10 keV flux without correcting Galactic absorption in a unit of 10$^{-11}$ erg s$^{-1}$ cm$^{-2}$.

  \footnotemark[$\|$] 2--10 keV luminosity with correcting Galactic absorption in a unit of 10$^{34}$ erg s$^{-1}$. The distance of 3.7 kpc is assumed.


 }\hss}}
\end{tabular}
\end{center}
\end{table}

\begin{table}
\begin{center}
\caption{The same as table~\ref{tab2} but with the model~2 (APEC).}
\label{tab3}
\begin{tabular}{llcc} \hline \hline
& & \multicolumn{2}{c}{APEC} \\
\cline{3-4}
Model & Parameter & Abundance = Free & 0.3 (fix) \\ \hline
phabs & $N_{\rm H}$$^{\rm Gal}$ ($\times10^{22}$cm$^{-2}$) & 0.49$^{+0.04}_{-0.02}$ & 0.38$\pm$0.02 \\
APEC & $kT$ (keV) & 1.75$^{+1.38}_{-0.63}$ & 39$^{+6}_{-8}$\\
 & Abundance\footnotemark[$*$] & 0 ($<$ 0.004) & 0.3 (fix) \\
 & Normalization & 1.54$^{+0.65}_{-0.30}$$\times$10$^{-2}$ & 1.56$^{+0.04}_{-0.09}$$\times$10$^{-2}$\\
compPS(seed=BB) &$T_{\rm BB}$ (keV) & 1.05$\pm$0.02 & 1.06$\pm$0.01\\
 &$\tau_{\rm BB}$ & 0.6$^{+0.04}_{-0.02}$ & $<$ 0.1\\
 &Normalization$_{\rm BB}$ & 3.22$^{+0.52}_{-0.27}$ & 1.94$^{+0.13}_{-0.05}$\\
 &$R_{\rm BB}$ (m) & 664$^{+52}_{-29}$ & 515$^{+19}_{-5}$\\ \hline
 &$\chi^{2}_{\nu}$(d.o.f) & 1.14 (587) & 1.15 (588)\\
 & $F_{2 - 10}$ & 4.08 & 4.09 \\
 & $L_{2 -10}$ & 6.98 & 6.93 \\ \hline

\multicolumn{3}{@{}l@{}}{\hbox to 0pt{\parbox{105mm}{\footnotesize
  \par\noindent

  \footnotemark[$*$] Abundance respect to the solar value of \citet{anders_abund}.

 }\hss}}
\end{tabular}
\end{center}
\end{table}

\begin{table}
\begin{center}
\caption{The same as table~\ref{tab2} but with the model~3.}
\label{tab4}
\begin{tabular}{llc} \hline \hline

Model & Parameter &  \\ \hline
phabs & $N_{\rm H}$$^{\rm Gal}$ ($\times10^{22}$cm$^{-2}$) & 0.65$\pm$0.15 \\
compPS(seed=DBB) & $T_{\rm DBB}$ (keV) & 0.11$\pm$0.02 \\
 &$\tau_{\rm DBB}$ & 0.2 ($\leq$ 0.4) \\
 &Normalization$_{\rm DBB}$\footnotemark[$*$] & 2.07$^{+5.13}_{-1.62}\times 10^4$ \\
 &$R_{\rm DBB}$ (km) & 75$^{+65}_{-40}$ \\
compPS(seed=BB) &$T_{\rm BB}$ (keV) & 1.05$\pm$0.05 \\
 &$\tau_{\rm BB}$ & 0.6$\pm$0.1  \\
 &Normalization$_{\rm BB}$ & 3.36$^{+0.34}_{-0.36}$ \\
 &$R_{\rm BB}$ (m) & 680$^{+30}_{-40}$ \\ \hline
 &$\chi^{2}_{\nu}$(d.o.f) & 1.14 (587) \\ 
 & $F_{2 - 10}$ & 4.09 \\
 & $L_{2 -10}$ & 7.11 \\ \hline

\multicolumn{3}{@{}l@{}}{\hbox to 0pt{\parbox{85mm}{\footnotesize
  \par\noindent
  \footnotemark[$*$] Normalization = ($R_{\rm DBB}$ (km))$^2$ / (0.37
 (10 kpc))$^2 {\rm cos}\theta$, where $R_{\rm DBB}$ is an inner radius of
the disk seed-photon emission in a unit of km. $\theta$ is an inclination angle and 60$^{\circ}$ is assumed.

 }\hss}}
\end{tabular}
\end{center}
\end{table}

\subsection{Analysis of Time Variation}\label{sec:ana_time}

\begin{figure}[htbp]

  \begin{center}
    \FigureFile(80mm,80mm){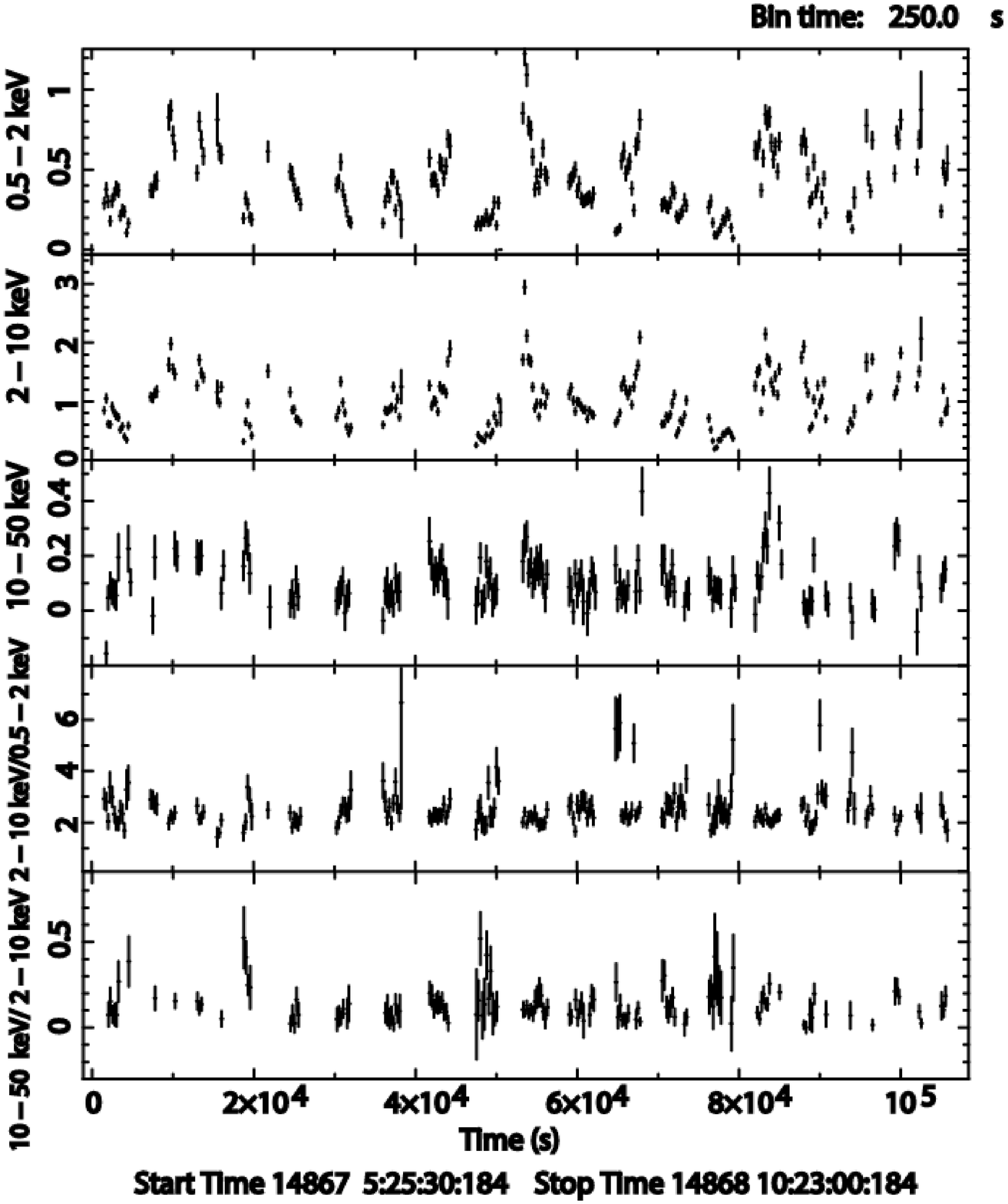}
  \end{center}
  \caption{X-ray light curves of XIS-0 0.5--2 keV (top row), 2--10 keV (2nd) and HXD-PIN 10--50 keV (3rd). Their hardness ratios between (2--10 keV)/(0.5--2 keV) and (10--50 keV)/(2--10 keV) are shown in 4th and 5th rows.}
\label{fig4}
\end{figure}

\begin{figure}[htbp]
 \begin{center}
    \FigureFile(80mm,80mm){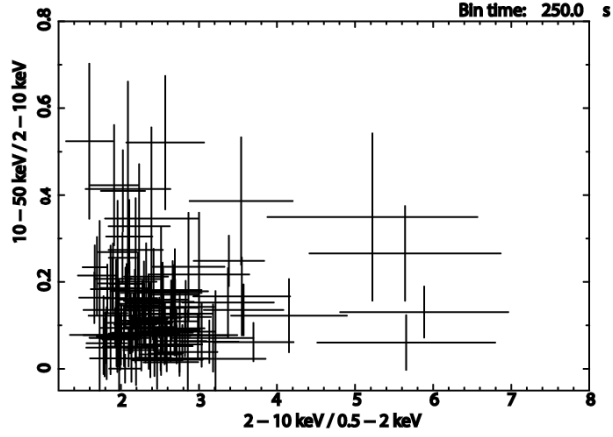}
 \end{center}
  \caption{Color-color diagram for the XIS-0 0.5--2 keV, 2--10 keV and
 HXD--PIN 10--50 keV energy bands.}
\label{fig5}
\end{figure}

We made the source light curve by using FTOOLS lcurve 1.0 and studied the time variation.
Figure~\ref{fig4} shows light curves of XIS-0 0.5--2 keV, 2--10 keV and HXD-PIN 10--50 keV,
and their color-color diagram is plotted in figure~\ref{fig5},
where the bin size is 250 s. 
The count rate varies by a factor of 1.5--4 with a time scale of 
$\sim$ few 100 s for 0.5--2 keV and 2--10 keV as also reported by \citet{masetti_syxb2}.

\begin{figure}[htbp]
 \begin{center}
    \FigureFile(80mm,80mm){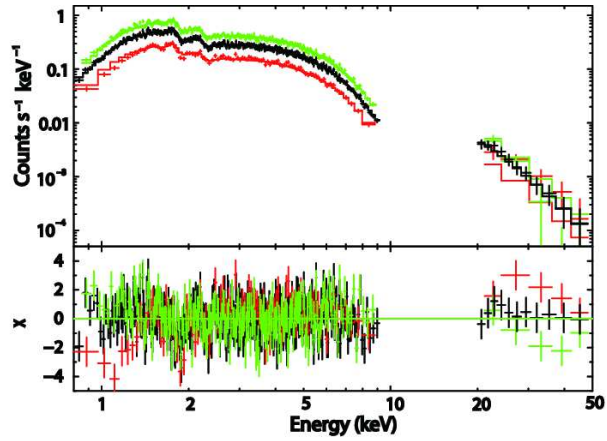}
 \end{center}
  \caption{(Top) time average spectra (black) and the spectra of the "high"- (green) and "low"- (red) luminosity periods.
(Bottom) residuals to the model~3 with only the total normalization left free.
}
\label{fig6}
\end{figure}

In order to investigate whether the color also changes with
the source intensity,
we defined a "high"-luminosity period as $> 1.65$ counts
s$^{-1}$ and a "low"-luminosity period as $< 1.05$ counts
s$^{-1}$ in the XIS-0 0.5--10 keV band.
Then, we made spectra of XIS-0, 1, 3 and HXD-PIN for both of the "high"- and "low"-luminosity periods.
To examine the change of these spectra, we fitted the spectra with the model~2 (BB).
First, we fixed the spectral parameters
representing the spectral shape to the best-fit parameters in section~\ref{sec:ana_ave},
and fitted the spectra with only the total normalization left free.
As a result shown in figure~\ref{fig6},
the normalization factors of the "high" and "low" spectra are obtained as 1.49$\pm$0.01 and 0.54$\pm$0.01 with the $\chi^2_{\nu}$ values of 1.31 (d.o.f. = 341) and 1.96 (d.o.f. = 146), respectively.
The "low" spectra have the worse $\chi^2_{\nu}$ value due to the relatively lower flux in the soft energy band.
We next fitted the spectra with setting the normalizations of 
the soft and hard components of the model~2 (BB)
 as the free parameters.
The obtained normalization values and their ratio 
are summarized in table~\ref{tab5-1}.
The same analysis with the model~3 is shown in table~\ref{tab5-2}.
Since the "low" spectra show the flux of the soft component at a lower level than that of the hard compPS (seed=BB)
 (i.e., the lower ratio by 40--50\%)
in both models,
it is indicated that the spectrum becomes harder when the source flux decreases.

\begin{figure}[htbp]
 \begin{center}
    \FigureFile(80mm,80mm){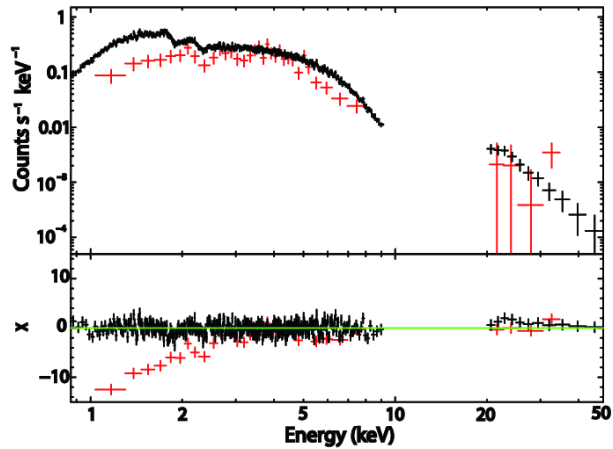}
 \end{center}
  \caption{The same as figure~6 but for the time-average spectra (black) and those of the "hard-color" period (red).}
\label{fig7}
\end{figure}

In figure~\ref{fig5}, there are 4 data points with a high count ratio ($>5$) between
2--10 keV and 0.5--2 keV, namely the color becomes hard.
These correspond to the period around the time of 65 ks in figure~\ref{fig4}, where
the XIS count rate is low in both energy bands.
We created the spectra for this period, and compared them with the time-average spectra in figure~\ref{fig7}.
Since the difference between them appears only below 2 keV, and the "hard-color" spectra cannot be represented by changing only the total normalization of the model~2 (BB) ($\chi^{2}_{\nu}$ = 3.80 (d.o.f = 101)).
We again fitted the spectra with changing the normalizations of
 the model~2 (BB) or 3, and found that the soft component 
is not required
 (see table~\ref{tab5-1} and \ref{tab5-2}).
Compared with the model~3, the soft component of the model~2 (BB) has the smaller contribution below 2 keV and cannot reproduce the spectra even with the normalization of zero.

This feature below 2 keV was represented better by the additional absorption $N_{\rm H}$ to the time-average spectra.
The model is "pcfabs$\times$phabs(BB + compPS (seed=BB))",
where "pcfabs" has the covering fraction from 0 to 1.
Then, as shown in table~\ref{tab5-3},
$N_{\rm H}$ is obtained as $\sim 4 \times 10^{22}$ cm$^{-2}$ and about one order of magnitude larger than that of the "phabs" value ($0.33$ and $0.65 \times 10^{22}$ cm$^{-2}$ for the model~2 (BB) and 3) with the large covering fraction $\sim 0.75$.
Therefore, there is a possibility that some material additionally covered the source in this period $\sim$ 1000 s.

%
%
%
%

\begin{table}
\begin{center}
\caption{Normalizations of the model~2 (BB) for spectra of "high"-luminosity, "low"-luminosity and "hard-color" periods.
}
\label{tab5-1}
\begin{tabular}{lcccc} \hline \hline

Parameter & High luminosity & Low luminosity & Hard color & Average\footnotemark[$\dagger$] \\ \hline
Normalization$_{\rm BB}$ & $(7.3\pm0.3) \times 10^{-5}$ & $(1.4\pm0.2) \times 10^{-5}$ & 0 & $4.2 \times 10^{-5}$ \\
Normalization$_{\rm BB}$ & 5.73$\pm$0.04 & 2.22$\pm$0.03 & 2.32 & 3.9 \\
Ratio\footnotemark[$*$] & $(1.27\pm0.06) \times 10^{-5}$ & $(0.63\pm0.10) \times 10^{-5}$ & 0 & $1.1 \times 10^{-5}$ \\ \hline
$\chi^{2}_{\nu}$ (d.o.f) & 1.26 (340) & 1.28 (145) & 2.42 (100) &   \\
$F_{0.5 - 2}$\footnotemark[$\ddagger$] & 0.83 & 0.28 & $\cdots$\footnotemark[$\S$] & 0.55 \\
$F_{2 - 10}$ \footnotemark[$\ddagger$]& 6.00 & 2.33 & $\cdots$\footnotemark[$\S$] &  4.08 \\ \hline
\multicolumn{5}{@{}l@{}}{\hbox to 0pt{\parbox{130mm}{\footnotesize
  \par\noindent
  \footnotemark[$*$] Ratio between two Normalization$_{\rm BB}$ parameters of the model~2 of "phabs(BB + compPS (seed=BB))".

  \footnotemark[$\dagger$] The time-average values of the model~2 in table~\ref{tab2}.

  \footnotemark[$\ddagger$] Flux without correcting Galactic absorption in a unit of 10$^{-11}$ erg s$^{-1}$ cm$^{-2}$.

  \footnotemark[$\S$] Not obtained due to the worse $\chi^{2}_{\nu}$.
 }\hss}}
\end{tabular}
\end{center}
\end{table}

\begin{table}
\begin{center}
\caption{The same as table~\ref{tab5-1} but for the model~3.
}
\label{tab5-2}
\begin{tabular}{lcccc} \hline \hline

Parameter & High luminosity & Low luminosity & Hard color & Average\footnotemark[$\dagger$] \\ \hline
Normalization$_{\rm DBB}$ & $(3.27\pm0.07) \times 10^{4}$ & $(0.90\pm0.04) \times 10^{4}$ & 0 ($<$270) & $2.07 \times 10^{4}$ \\
Normalization$_{\rm BB}$ & 5.06$\pm$0.05 & 2.04$\pm$0.03 & 2.62$\pm$0.12 & 3.36 \\
Ratio\footnotemark[$*$] & $6460\pm210$ & $4370\pm280$ & 0 ($<$110) & 5970 \\ \hline
$\chi^{2}_{\nu}$ (d.o.f) & 1.26 (340) & 1.30 (145) & 1.56 (100) &   \\
$F_{0.5 - 2}$\footnotemark[$\ddagger$] & 0.84 & 0.28 & 0.19 &  0.55 \\
$F_{2 - 10}$ \footnotemark[$\ddagger$]& 6.02 & 2.32 & 2.70 &  4.09 \\ \hline
\multicolumn{5}{@{}l@{}}{\hbox to 0pt{\parbox{130mm}{\footnotesize
  \par\noindent
  \footnotemark[$*$] Ratio between Normalization$_{\rm DBB}$ and Normalization$_{\rm BB}$ of the model~3 of "phabs(compPS (seed=DBB) + compPS (seed=BB))".

  \footnotemark[$\dagger$] The time-average values of the model~3 in table~\ref{tab4}.

  \footnotemark[$\ddagger$] Flux without correcting Galactic absorption in a unit of 10$^{-11}$ erg s$^{-1}$ cm$^{-2}$.

 }\hss}}
\end{tabular}
\end{center}
\end{table}

\begin{table}
\begin{center}
\caption{Parameters of the additional absorber ("pcfabs") observed in the "hard-color" spectra\footnotemark[$*$].
}
\label{tab5-3}
\begin{tabular}{lc} \hline \hline

Parameter \\ \hline
$N_{\rm H}$ ($\times 10^{22}$ cm$^{-2}$) & $3.8^{+0.9}_{-0.7}$ \\
Covering fraction & $0.75^{+0.05}_{-0.04}$ \\ \hline
$\chi^{2}_{\nu}$ (d.o.f) & 1.06 (100) \\ \hline
\multicolumn{2}{@{}l@{}}{\hbox to 0pt{\parbox{80mm}{\footnotesize
  \par\noindent
  \footnotemark[$*$] The continuum emission of "phabs(BB + compPS (seed=BB))" is fixed at the time-average values of the model~2 (BB) in table~\ref{tab2}.

 }\hss}}
\end{tabular}
\end{center}
\end{table}

%
%

\begin{figure}[htbp]
 \begin{center}
    \FigureFile(80mm,80mm){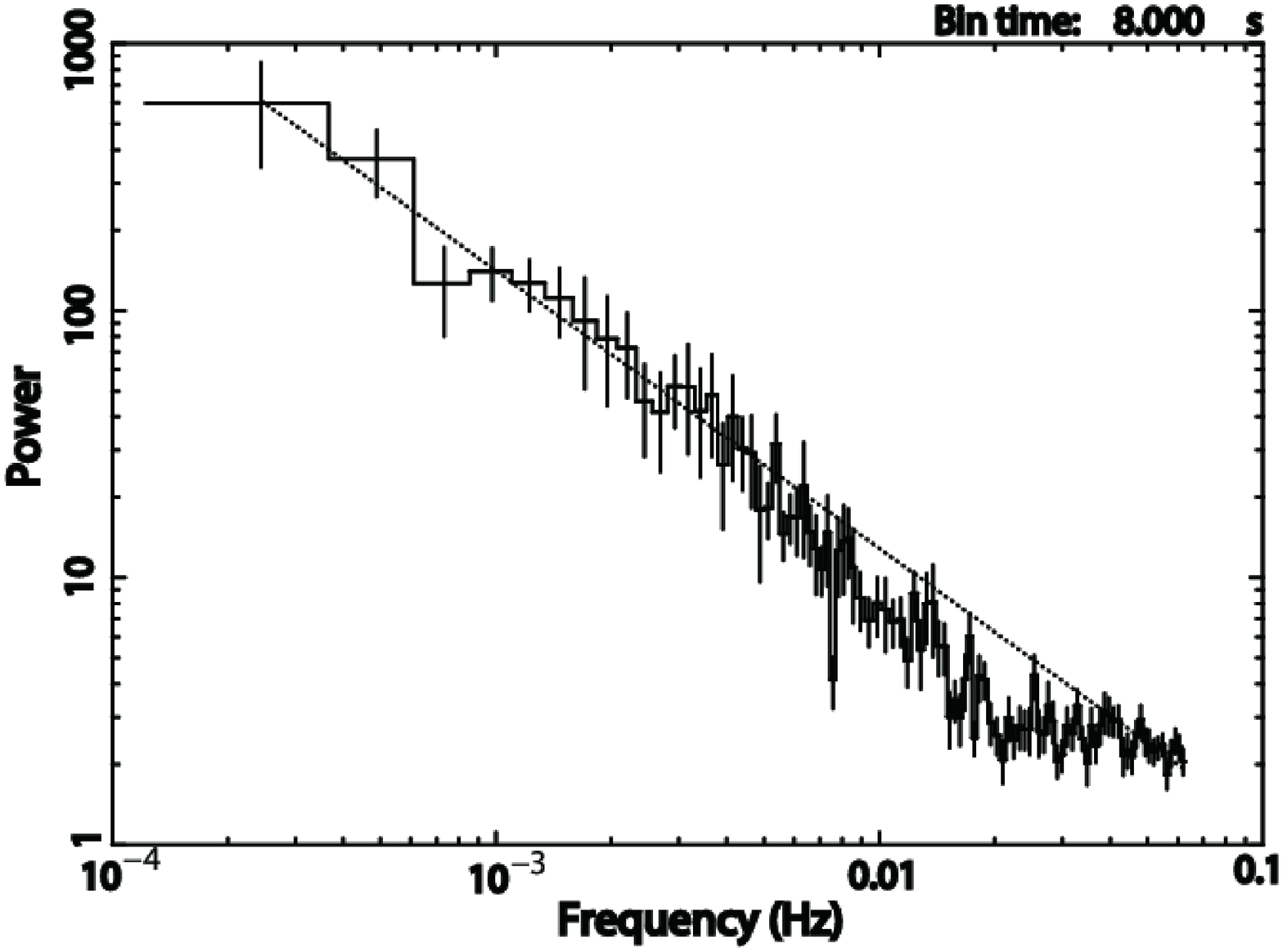}
    \FigureFile(80mm,80mm){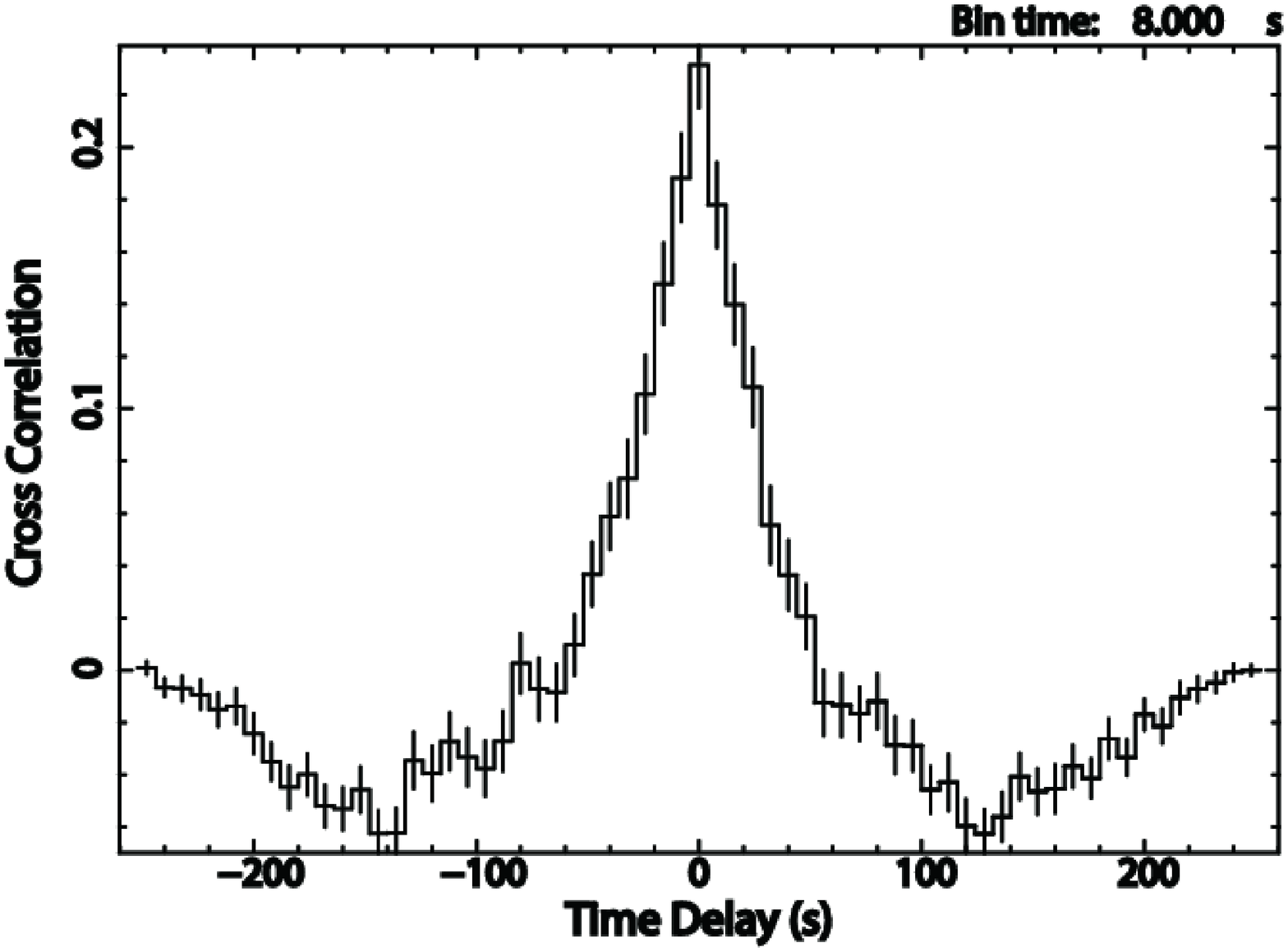}
 \end{center}
  \caption{(Left) a power spectrum of IGR J16194--2810 from the XIS light curve.
A dot line shows the best-fit power-law model of $1/f^{1.04 \pm 0.02}$.
(Right) a cross-correlation function calculated from the XIS light curve between 0.8--1.5 keV and 1.5--10 keV energy bands.
}
\label{fig8}
\end{figure}

Since some SyXBs exhibit an X-ray pulse (e.g., GX 1+4, 4U 1954+31), we searched the X-ray pulse for IGR J16194--2810.
Figure~\ref{fig8} shows the power spectrum created by FTOOLS powspec 1.0 for the XIS light curve,
where the XIS0, XIS1 and XIS3 light curves of the source region were combined into one to increase the statistic and were performed the barycentric correction by aebarycen of FTOOLS.
The minimum time resolution is 8 s determined by the XIS exposure period,
and we calculated the power spectrum using 4096-s data segments and averaged them.

In figure~\ref{fig8}, the power is proportion to $1/f^{1.04 \pm 0.02}$,
where $f$ is the frequency, and resembles to "very-low-frequency noise" observed in LMXBs \citep{hasinger_lmxb}.
There are no significant periodic features in the time scale from 0.0002 to 0.06 Hz (i.e., from 16 to 4096 s)
in the same way as the previous report \citep{masetti_syxb2}.
Since the total number of the X-ray events in the XIS light curve is $N_{\rm ph}$ = 212806,
the upper limit of the amplitude $A_{\rm UL}$ of the sinusoidal modulation is $A_{\rm UL} \sim (2.6 Power / (0.773 N_{\rm ph}))^{1/2}$ \citep{vanderklis_timing},
resulting in $A_{\rm UL} \sim$ a few to 15\% over 0.06 Hz to 0.0002 Hz.

If there is any lag between light curves of the soft and hard energy bands,
such information is useful to identify the origin of the soft emission component from
whether the NS surface (BB) or the accretion disk (compPS (seed=DBB)),
since the disk region presumably becomes brighter before the NS surface.
Then, using all XIS events in the source region by FTOOLS crosscor 1.0, we calculated the cross correlation of the light curves in 0.8--1.5 keV and 1.5--10 keV with the time resolution of 8 s,
where the data were divided into 256-s segments and averaged.
However, as shown in figure~\ref{fig8}, the peak is consistent at the 0-s delay and has a symmetric shape, and no significant lags are detected between them.

\section{Discussion}\label{sec:discussion}

We observed the SyXB IGR J16194--2810 in the low/hard state with Suzaku.
The source signal was detected in 0.8--50 keV wide-energy band simultaneously, and the luminosity was $L \sim 7\times10^{34}$ erg s$^{-1}$ in 2--10 keV.
The X-ray spectrum can be described by the model~2 (BB) "phabs(BB + compPS (seed=BB))" or the model~3 "phabs(compPS (seed=DBB) + compPS (seed=BB))".
The hard compPS (seed=BB) component represents the energy band above 2 keV and gives the seed-photon parameters of $T_{\rm BB} \sim 1.0$ keV, $\tau_{\rm BB} \sim 0.6$ and $R_{\rm BB} \sim 700$ m.
The soft energy band is reproduced by either the raw BB ($T_{\rm BB} \sim$ 0.4 keV, $R_{\rm BB} \sim$ 1.7 km) or compPS (seed=DBB) ($T_{\rm DBB} \sim$ 0.1 keV, $\tau_{\rm DBB} \sim 0.1$, $R_{\rm DBB} \sim$ 75 km
 but with large uncertainties).
During the net exposure of 46 ks, the flux varied by a factor of 4 in the XIS band,
and the spectral shape becomes harder when the flux decreased.
There was also a time when the count rate only below 2 keV decreased for $\sim$ 1000 s (the "hard-color" period).
The power spectrum of the light curve shows the power-law shape
without any significant periodic features.

As described in \S\ref{sec:ana_ave}, $R_{\rm BB}$ and $T_{\rm BB}$
of the hard compPS (seed=BB) component are well
constrained by the spectral analysis, and thus we discuss these parameters.
We obtained $R_{\rm BB}$ of IGR J16194--2810 to be $\sim$ 700 m.
This value is smaller than the NS radius of $\sim$ 10 km, indicating that the BB emission comes from a part of the NS surface.
We compare this issue with an SyXB 4U 1700+24 and typical 
transient and persistent
LMXBs,
Aql X-1 and 4U 0614+091 \citep{sakurai_aqlx1,singh_4u0614}.
Table~\ref{tab6} summarizes the X-ray luminosity, $T_{\rm BB}$, and $R_{\rm BB}$ of these objects together with those of IGR J16194--2810.
Since the parameters of the hard compPS (seed=BB) component are almost independent from the modeling, we listed those of the model~3.
Note that the spectrum of Aql X-1 in the high/soft state, that in the low/hard state and that of 4U 0614+091 in the low/hard state are fitted with "diskbb + nthcomp", "diskbb + compPS (seed=BB)", and "BB + compST", respectively,
where the nthcomp and compST models also calculate the Comptonized emission by
\citet{zycki_nthcomp} and \citet{sunyaev_compst}.
In the case of 4U 0614+091, the Comptonized emission of compST is separately treated from the BB model, and the BB radius is considered as the lower limit without including the contribution of the seed photons.
Considering the spectral features of the SyXBs and LMXBs,
we suggested that the hard energy band of X-ray spectra of the NS-LMXB in the low/hard state can be explained commonly by
the hard Comptonized BB emission from the NS surface.

\begin{table}
\begin{center}
\caption{Comparison of the X-ray luminosity, $T_{\rm BB}$ and $R_{\rm BB}$.}
\label{tab6}
\begin{tabular}{lccccc} \hline \hline

 & \multicolumn{2}{c}{Aql X-1} & 4U 0614+091 & IGR J16194--2810 & 4U 1700+24 \\ \cline{2-3}
Parameter{\textbackslash}State & high/soft & low/hard & low/hard & low/hard & low/hard \\ \hline
$T_{\rm BB}$ (keV) & 1.4$^{+0.2}_{-0.1}$ & 0.51$\pm$0.02 & $\simeq$ 0.6 & 1.05 $\pm$ 0.05 & 0.76--1.07 \\
$R_{\rm BB}$ (km) \footnotemark[$*$] & 3.0$^{+0.8}_{-0.7}$ & 10 $\pm$ 2 & 2--6 & 0.68$^{+0.03}_{-0.04}$ & 0.045--0.24 \\
$L$ ($\times 10^{36}$erg s$^{-1}$) \footnotemark[$*$] &  15 & 2.5 & 3--4 & 7$\times10^{-2}$ & 2$\times10^{-4} - 2\times10^{-2}$ \\
References & [1] & [1] & [2] & [3] & [4] \\ \hline
\multicolumn{6}{@{}l@{}}{\hbox to 0pt{\parbox{150mm}{\footnotesize
  \par\noindent
  \footnotemark[$*$] We assume a distance of objects as
 5.2 kpc for Aql X-1 \citep{jonker_aqlx1},
 5.0 kpc for 4U 0614+091 \citep{swank_4u0614},
 3.7 kpc for IGR J16194--2810 \citep{masetti_syxb2} and
 420 pc for 4U 1700+24 \citep{masetti_4u1700}.

  \footnotemark[] References [1] \citet{sakurai_aqlx1} [2] \citet{singh_4u0614}  [3] This work [4] \citet{nagae_4u1700}

 }\hss}}

\end{tabular}
\end{center}
\end{table}

From table~\ref{tab6}, it can be seen that $R_{\rm BB}$ of IGR J16194--2810 and
4U 1700+24 is smaller than 1 km, while that of Aql X-1 and 4U 0614+091
is larger and consistent with that 
the emission comes from most of the NS surface.
The temperature value $T_{\rm BB}$ of the SyXBs is $\sim$ 1 keV and higher than that of the LMXBs,
although that of Aql X-1 in the high/soft state is even higher.
Therefore, to irradiate the observed luminosity in the low/hard state,
the SyXBs have the smaller radiation region on the NS surface with the higher seed-photon temperature,
compared with the LMXBs.
We suggest that the difference of luminosity-dependence of $R_{\rm BB}$ and $T_{\rm BB}$ is due to the different magnetic field strength $B$ between LMXBs and SyXBs.

We speculate that the magnetic field of SyXBs is stronger than that of LMXBs, and thus the matter from the donor star is caught by the magnetic field and accreted to both poles along a line of magnetic force when the accreted rate is low. 
This hypothesis is also described in \citet{masetti_syxb2}.
For LMXBs, the magnetic field is generally believed to be weak, and thus the accreting stream falls onto the NS boundary layer without effects of the magnetic field.

In order to discuss this possibility, it is important to consider
 the Alfven radius $R_{\rm A} = 2.9 \times 10^{8} M^{1/7}_1 R^{-2/7}_{6} L^{-2/7}_{37} \mu^{4/7}_{30}$ cm,
where $M_{1}$ is a mass of the compact star with the unit of 1\MO, $R_{6}$ is the radius of the compact star with a unit of 10$^6$ cm, $L_{37}$ is the luminosity with a unit of 10$^{37}$ erg s$^{-1}$, and $\mu_{30}$ is a magnetic moment of the compact star with the units 10$^{30}$ G cm$^{3}$. 
An NS with $B$ $\cong$ 10$^{12}$ G, $R$ $\cong$ 10$^6$ cm has $\mu_{30}$ $\cong$ 1 \citep{frank_alfven}.
When $L$ = 10$^{34}$, 10$^{36}$ and 10$^{38}$ erg s$^{-1}$, 
$R_{\rm A}$ becomes 30 km, 8 km, and 2 km, respectively, for $B$ $\sim$
10$^{7}$ G, and 400 km, 110 km and 30 km, respectively, for $B$ $\sim$ 10$^{9}$ G.
Within the Alfven radius, the accreting matter flows along a magnetic field line and finally falls on the magnetic poles of the compact star.
In the case of a low luminosity and/or a strong magnetic field of SyXBs, $R_{\rm A}$ could be larger than the NS radius and the accreting stream falls on the magnetic pole region which is a small area, leading  to a small $R_{\rm BB}$.
The accreting matter is shocked on the NS surface and then heated up to $\sim$ 1 keV.
In the case of a high luminosity and/or a weak magnetic field of typical LMXBs, $R_{\rm A}$ is similar or smaller than the NS radius and the accreting matter falls on a large portion of the NS surface.
As a result of the Stefan-Boltzmann law, the temperature $T_{\rm BB}$ is not so high as 1 keV. 

\citet{lamb_hotspot} estimated the size and raidus of the hot spot, where the matter accrets on the NS surface and emits the X-rays, as
$\pi R^2(R/R_{\rm A})$ and $R\sqrt{(R/R_{\rm A})}$, respectively.
The value of $R$ is the NS radius of 10 km.
When we assume that the observed $R_{\rm BB}$ of $\sim$ 700 m corresponds to the hot-spot radius, $R_{\rm A}$ is calculated as $\sim$ 2000 km and $B$ becomes $\sim 4 \times 10^{10}$ G with the luminosity of $7 \times 10^{34}$ erg s$^{-1}$.
The strength of the magnetic field is actually higher than that of typical LMXBs $< 10^9$ G.

If the accreting matter falls on the magnetic poles, the X-ray emission should show pulsation as HMXBs.
However, we could not detect the pulse from IGR J16194--2810, and also 4U 1700+24 (Nagae, private communication).
\citet{lu_syxb}
 listed 10 SyXBs and candidates.
The X-ray pulse has been detected from several objects; for example, GX 1+4 and 4U 1954+31.
Their spin was $P_{\rm s}$ $\sim$ a few 100 s
(e.g., \cite{chakrabarty_gx1+4}, \cite{gonzalez_gx1+4})
 and $\sim$ 5 h \citep{corbet_4u1954},
 which is longer than typical spin periods of LMXBs.
Therefore, we speculate that the spin period of IGR J16194--2810 may be too long
 ($\gtrsim$ 1 h)
 to be detected during this observation.
There are also other possibilities to explain the absence of the pulsation that the spin axis just coincides with the magnetic one,
it can be wiped out through Comptonization by a very hot Compton cloud \citep{torrejon_hard},
or the compact star might be a black hole without a solid surface.

To accrete the matter smoothly from $R_{\rm A}$ to the NS surface,
the NS needs to spin slower (i.e., have a longer spin period) than
the equilibrium spin period $P_{\rm eq} \geq 2\pi$$\sqrt[]{R_{\rm A}^{3}/GM}$,
where $G$ is a gravitational constant and $M$ is an NS mass.
Otherwise, the matter is likely to be ejected due to the "propeller" effect
(e.g., \cite{illarionov_propeller}).
Assuming $P_{\rm eq}$ = 1 h
 from the above discussion, $R_{\rm A}$ becomes
 $\leq 4 \times 10^{5}$ km.
Then, the magnetic field is estimated as
 $B \leq 4 \times$ 10$^{14}$ G
 with the observed luminosity $L$ = 7 $\times$ 10$^{34}$ erg s$^{-1}$.
This upper limit is as large as that of magnetars and
also consistent with the value estimated above from the radius of the hot spot.

From the analysis of the time-average spectrum, the origin of the soft energy component below 2 keV is considered as the emission from the NS surface (mode~2 (BB)) or the accreting stream (model~3).
The timing analysis does not show significant lags and it is difficult to constrain the geometry furthermore.
If the emission comes from the NS surface, the region of $\sim 1.7$ km is larger than that of the hard component ($\sim$ 700 m) but still smaller than the NS radius.
Additionally, the temperature of $\sim 0.4$ keV is lower than the hard one $\sim 1.0$ keV,
and this component may arise from the thermalized matter after accreted at the magnetic poles.
On the other hands, there are also several predictions about the accreting matter onto magnetic compact objects whether it forms an accretion disk or a shell structure
\citep{mitsumoto_disk,shakura_shell}.
However, the disk inner radius $R_{\rm DBB}$ of the model~3 is obtained as 75$^{+65}_{-40}$ km and one or two orders of magnitude smaller than the estimated Alfven radius $R_{\rm A} \sim$ 2000 km.
This might imply that there exists a disk or shell structure around $R_{\rm A}$
but $R_{\rm DBB}$ obtained by the spectral fitting is underestimated.
In fact, we cannot constrain $R_{\rm DBB}$ and $T_{\rm DBB}$ well
or detect any large fraction of the un-Comptonized raw emission
(i.e., small portion of the emission might be injected into the Compton cloud and observed)
due to the interstellar absorption as described in \S\ref{sec:ana_ave}.
In terms of the incompatibility between $R_{\rm DBB}$ and $R_{\rm A}$,
the model~2 (BB) reproduces the data physically better than the model~3.


During the "hard-color" period of $\sim$ 1000 s,  
the source flux decreased only below 2 keV.
The spectrum is well represented by the additional absorber with
the column density of $\sim 4 \times 10^{22}$ cm$^{-2}$
and the covering fraction of $\sim 0.75$.
Such a behavior is also observed in other X-ray binaries
due to a clumpy stellar wind
 (e.g., dips in Cyg X-1; \cite{feng_cygx1dip}).
The red giant of 
IGR J16194--2810
might also have the clumpy stellar wind
\citep{crowley_clump}.




\section{Summary}\label{sec:summary}

In this paper, we analyzed the Suzaku data of SyXB IGR J16194--2810 in the energy range of 0.8--50 keV, and obtained the results as below.

\begin{enumerate}
\item
The time-average spectrum at the luminosity of $\sim 7 \times 10^{34}$ erg s$^{-1}$ was physically represented by
the model~2 (BB) (blackbody plus Comptonization model) or the model~3 (two-Comptonizations model).
In both models, there were two emission components;
the hard component of a Comptonized BB emission, and the soft one of either a raw BB or a Comptonized emission.
\item
Compared with other SyXB and typical transient and persistent LMXBs in the low/hard state, the hard emission component of the SyXBs has the smaller $R_{\rm BB} <$ 1 km and the higher $T_{\rm BB} \sim$ 1 keV.
We propose that this behavior is due to the stronger magnetic field of SyXBs than that of the LMXBs,
and the accreting stream falls on the magnetic poles of the NS.
\item
The emission region of the soft component is still unclear, since the lower energy band has uncertainties due to the interstellar absorption and there was no time lags observed between soft and hard energy bands.
One possibility (model~2 (BB)) is that the raw BB emission may arise from the thermalized matter after accreted at the magnetic poles.
The other (model~3) is that the seed photon of the Comptonized emission might be injected from small fraction of the accreting stream.
\item
The light curve showed time variation during the 1-d observation, and it was suggested that the spectrum becomes harder when the source flux decreases.
There was $\sim$ 1000-s timing when the flux only below 2 keV decreased.
The spectrum was reproduced by the additional absorber to the time-average one.
The power spectrum did not have any significant periodic features in the time scale from 0.0002 to 0.06 Hz.
\end{enumerate}

\end{document}